\documentclass{emulateapj}

\newcommand{\HI}{\ifmmode{\rm HI}\else{\mbox{H\,{\sc{i}}}}\fi}
\newcommand{\e}[1]{\ensuremath{\times 10^{#1}}}

\newcommand{\msun}{\ensuremath{{\rm \,M}_{\sun}}}
\newcommand{\msunns}{\ensuremath{{\rm M}_{\sun}}}
\newcommand{\Zsun}{\ensuremath{{\rm \,Z}_{\sun}}}

\newcommand{\kms}{\ensuremath{{\rm \,km\,s}^{-1}}}
\newcommand{\cm}{\ensuremath{{\rm \,cm}^{-2}}}
\newcommand{\Gyr}{\ensuremath{{\rm \,Gyr}}}
\newcommand{\kpc}{\ensuremath{{\rm \,kpc}}}
\newcommand{\pc}{\ensuremath{{\rm \,pc}}}
\newcommand{\kpcns}{\ensuremath{{\rm kpc}}}

\newcommand{\Jykms}{\ensuremath{{\rm \,Jy\,\kms{}}}}
\newcommand{\Jykmsns}{\ensuremath{{\rm Jy\,\kms{}}}}
\newcommand{\vlsr}{\ensuremath{{v}_{\rm LSR}}}
\newcommand{\rvir}{{\ensuremath{r_{\mathrm{vir}}}}}
\newcommand{\Mvir}{{\ensuremath{M_{\mathrm{vir}}}}}
\newcommand{\rmw}{\ensuremath{r_{\rm \,MW}}}
\newcommand{\lstar}{\ensuremath{L_\star}}
\newcommand{\simgt}{\ensuremath{\ga}}
\newcommand{\simlt}{\ensuremath{\la}}

\newcommand{\Fig}[1]{Figure~\ref{fig:#1}}

\newcommand{\Tab}[1]{Table~\ref{tab:#1}}
\newcommand{\Sec}[1]{Section~\ref{sec:#1}}

\newcommand{\codet}[1]{{\mbox{\sc#1}}}
\newcommand{\GCD}{{\codet{GCD+}}}

\newcommand{\pview}{{\codet{PView}}}
\newcommand{\cloudy}{{\codet{Cloudy}}}

\begin{document}

\title{
  On the origin of anomalous velocity clouds in the Milky Way.
}
\shorttitle{Origin of HVCs in the MW}

\author{
  Tim W. Connors\altaffilmark{1},
  Daisuke Kawata\altaffilmark{1,2},
  Jeremy Bailin\altaffilmark{1},
  Jason Tumlinson\altaffilmark{3},
  Brad K. Gibson\altaffilmark{4}
}

\altaffiltext{1}{Centre for Astrophysics \& Supercomputing, 
  Swinburne University, Hawthorn, VIC 3122, Australia}
\altaffiltext{2}{The Observatories of the Carnegie Institution of Washington,
  813 Santa Barbara Street, Pasadena, CA 91101, USA}
\altaffiltext{3}{Yale Center for Astronomy and Astrophysics,
  Department of Physics, P.O. Box 208101, New Haven, CT 06520, USA}
\altaffiltext{4}{Centre for Astrophysics, University of Central
  Lancashire, Preston, PR1 2HE, United Kingdom}
\shortauthors{Connors et al.}
\email{tconnors@astro.swin.edu.au}

\begin{abstract}
  We report that neutral hydrogen (\HI{}) gas clouds, resembling High
  Velocity Clouds (HVCs) observed in the Milky Way (MW), appear in
  MW-sized disk galaxies formed in high-resolution Lambda Cold Dark
  Matter ($\Lambda$CDM) cosmological simulations which include
  gas-dynamics, radiative cooling, star formation, supernova feedback,
  and metal enrichment.  Two such disk galaxies are analyzed, and
  \HI{} column density and velocity distributions in all-sky Aitoff
  projections are constructed.  The simulations demonstrate that
  $\Lambda$CDM is able to create galaxies with sufficient numbers of
  anomalous velocity gas clouds consistent with the HVCs observed
  within the MW, and that they are found within a galactocentric
  radius of $150 \kpc$.  We also find that one of the galaxies has a
  polar gas ring, with radius $30\kpc$, which appears as a large
  structure of HVCs in the Aitoff projection.  Such large structures
  may share an origin similar to extended HVCs observed in the MW,
  such as Complex~C.
\end{abstract}


%*****************************************************************
\section{Introduction}
\label{sec:intro}

High Velocity Clouds (HVCs) are neutral hydrogen (\HI{}) gas clouds
with velocities inconsistent with galactic rotation \citep{ww97}.
From our vantage point within the Galaxy, they appear to cover a large
portion of the sky relatively isotropically.  HVCs do not appear to
possess a stellar component \citep[e.g.][]{sb02,smg+05} and their
distances (and masses) are generally unknown.  Direct distance
constraints have only been made for a select number of HVCs
\citep{w01,tpg+06}.  There are still open questions as to whether HVCs
are local to the Milky Way (MW) or distributed throughout the Local
Group (LG); whether they are peculiar to the MW or are common in disk
galaxies; whether they are gravitationally bound or pressure confined;
whether they contain Dark Matter (DM); and their degree of
metal-enrichment.

\citet{pbg+04} report that there are no HVC-like objects with \HI{}
mass in excess of $4\e5\msun$ in three LG-analogs.  They suggest that
if HVCs are a generic feature, they must be clustered within $160\kpc$
of the host galaxy, ruling out the original \citet{bst+99} model in
which HVCs are gas clouds distributed in filaments on Mpc-scales.
\citet{tbw+04} and \citet{wbt+05} find 16 HVCs around M31, with \HI{}
masses ranging from $10^4$ to $6\e5 \msun$.  Most of the HVCs are at a
projected distance $< 15\kpc$ from the disk of M31.  Some clouds
appear to be gravitationally dominated by either DM or as yet
undetected ionized gas.  They also found two populations of clouds,
with some of the HVCs appearing to be part of a tidal stream, and
others appearing to be primordial DM dominated clouds, left over from
the formation of the LG.  Finally, it has also been speculated that
cooling instabilities in Cold Dark Matter (CDM) halos could lead to
clouds within $150\kpc$ of MW type galaxies, with HVC-like properties
\citep{mb04}.

In this {\it Letter}, we report that these mysterious \HI{} clouds
also appear in MW-size disk galaxies formed in $\Lambda$CDM
cosmological simulations.  We demonstrate that the simulated galaxies
show HVCs comparable in population to the observed ones.  We also find
that large HVCs resembling Complex~C appear in simulated galaxies.
Therefore, we conclude that HVCs appear to be a natural byproduct of
galaxy formation in the $\Lambda$CDM Universe.  The next section
describes our methodology, including a brief description of the
simulations and how we ``observe'' the simulated disk galaxies.  In
\Sec{results}, we show our results and discuss our findings.

\section{Methodology}
\label{sec:methodology}

We analyze two disk galaxy models found in cosmological simulations
that use the multi-mass technique to self-consistently model the
large-scale tidal field, while simulating the galactic disk at high
resolution. These simulations include self-consistently many of the
the important physical processes in galaxy formation, such as
self-gravity, hydrodynamics, radiative cooling, star formation,
supernova feedback, and metal enrichment. The disk galaxies we analyze
correspond to ``KGCD'' and ``AGCD'' in \citet{bkg+05}, and we use
these names hereafter.  Both simulations are carried out with our
galactic chemodynamics package \GCD{} \citep{kg03a}.

\begin{deluxetable*}{lrrrrrrrrrr}
  \tablecaption{
    Properties of simulations
  }
  \tablewidth{0pt}
  \tablehead{\colhead{Name} & \colhead{$\Mvir$}         &
    \colhead{$\rvir$}       & \colhead{$r_{\rm{disk}}$} &
    \colhead{$m_{\rm gas}$} & \colhead{$m_{\rm DM}$}    & 
    \colhead{$e_{\rm gas}$} & \colhead{$e_{\rm DM}$}    & 
    \colhead{$\Omega_0$}    & \colhead{$h_0$}           &
    \colhead{$\Omega_b$}    \\
    & \colhead{$(\msunns)$} & \colhead{$(\kpcns)$}      &
    \colhead{$(\kpcns)$}    & \colhead{$(\msunns)$}     &
    \colhead{$(\msunns)$}   & \colhead{$(\kpcns)$}      &
    \colhead{$(\kpcns)$}    & & & \\
  }
  \startdata
  KGCD & $8.8\e{11}$ & 240 & 10 & $9.2\e5$ & $6.2\e6$ & 0.57 & 1.1 & 0.3 &
  0.7 &  0.039 \\
  AGCD & $9.3\e{11}$ & 270 & 21 & $3.3\e6$ & $1.9\e7$ & 0.87 & 1.5 & 0.3 &
  0.65 & 0.045 \\
  \enddata
  \label{tab:params}
\end{deluxetable*}

The details of these simulations are given in \citet{bkg+05}.
\Tab{params} summarizes the simulation parameters and the properties
of the galaxies.  Column 1 is the galaxy name; Column 2, the virial
mass; Column 3, the virial radius; Column 4, the radial extent of the
gas disk, defined as the largest radius at which we find gas particles
in the disk plane.  Columns 5 and 6 contain the mass of each gas and
DM particle in the highest resolution region, and Columns 7 and 8 are
the softening lengths in that region.  The cosmological parameters are
presented in Columns 9--11.  Note that the spatial resolution for the
gas is determined by the smoothing length of the smoothed particle
hydrodynamics scheme.  The minimum smoothing length is set to be half
of the softening length of the gas particles (see \citealp{kg03a}).
The smoothing length depends on the density, and the average smoothing
length in the simulated HVCs we focus on here is $\sim 5 \kpc$.

\begin{figure}[t]
  \epsscale{1.1}
  \plotone{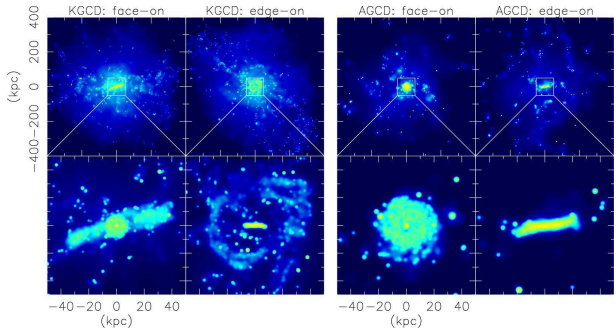}
  \caption[\HI{} gas column density maps]{\HI{} gas column density
    maps of KGCD (left set of 4 panels) and AGCD (right set)
    simulations.}
  \label{fig:gasdens}
\end{figure}

Both galaxies are similar in size and mass to the MW, and have clear
gas and stellar disk components.  \Fig{gasdens} shows edge-on and
face-on views of the projected gas density of each galaxy at the final
timestep.  We use the simulation output at $z=0.1$ for KGCD, as
contamination from low-resolution particles in the simulated galaxy
start to become significant at this redshift.  We use the output at
$z=0$ for AGCD.

\begin{figure}[t]
  \epsscale{1.1}
  \plotone{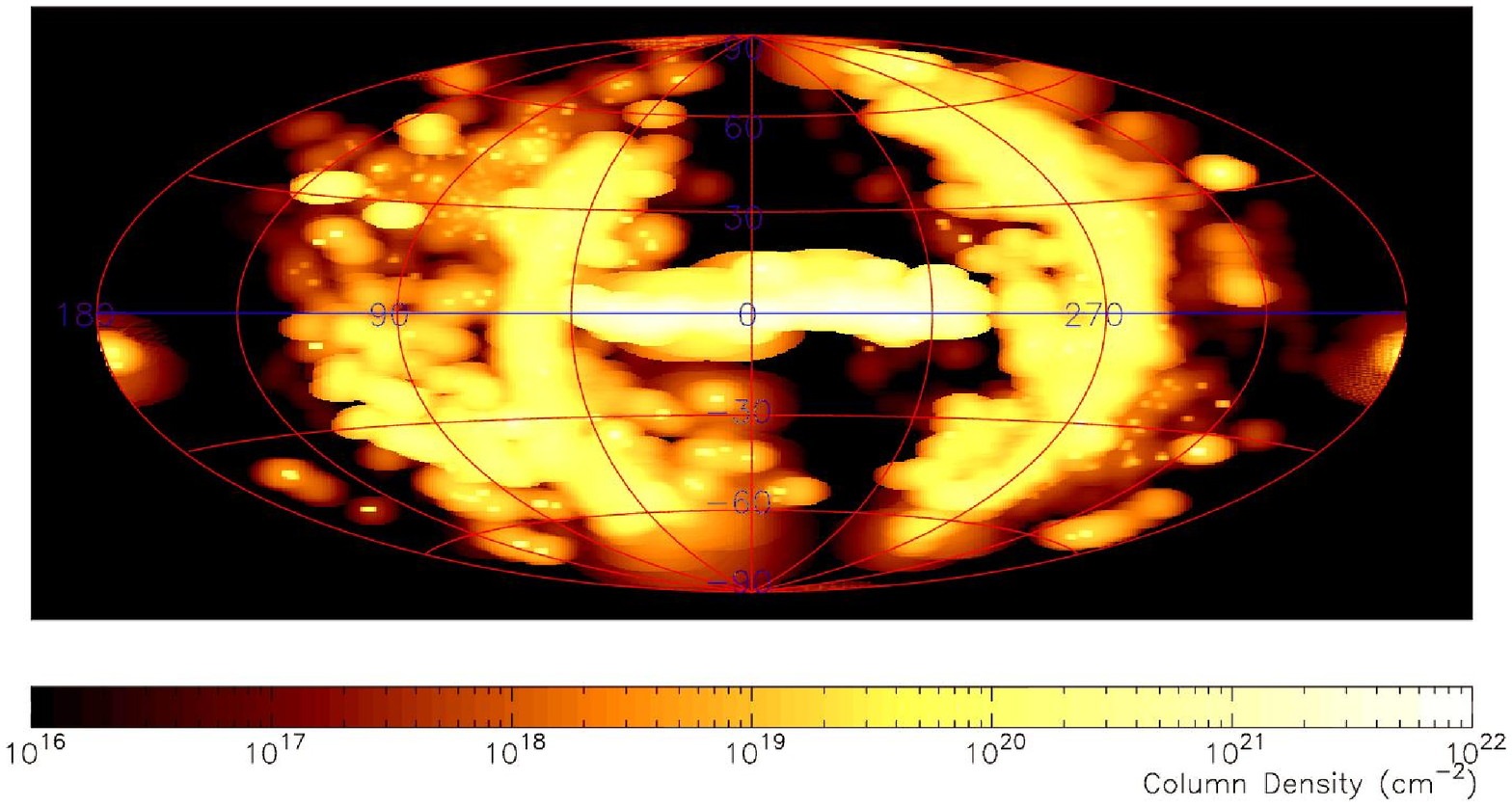}\\
  \plotone{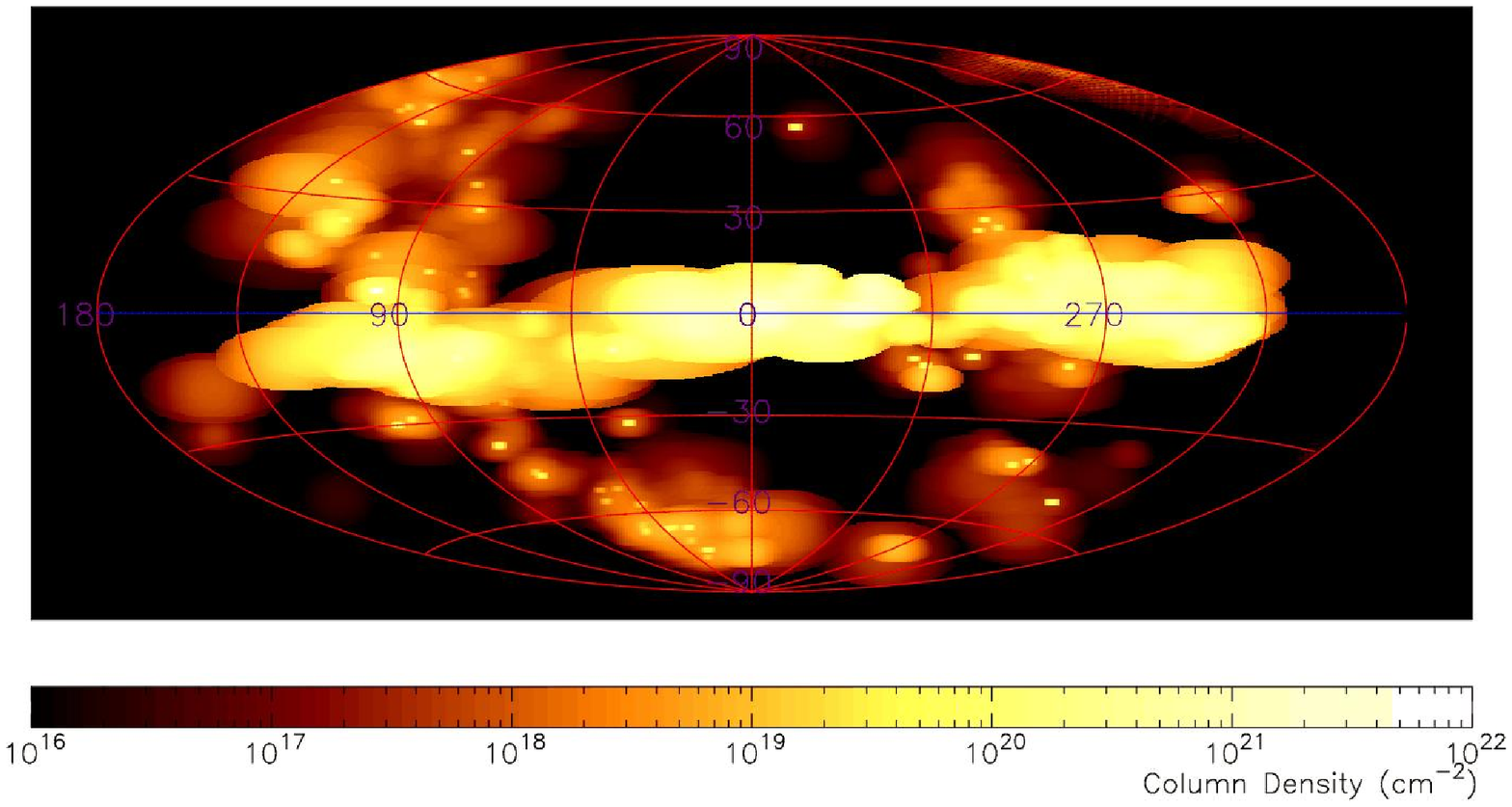}
  \caption[\HI{} column density maps]{\HI{} column density maps of KGCD
    (top) and AGCD (bottom) simulations, in a all-sky Aitoff
    projection in using galactic coordinates.  Particles with $|\vlsr|
    < 100\kms$ are excluded.  Compare these maps with fig.~2 in
    \citet{w91}.}
  \label{fig:HI_mom0}
\end{figure}

In order to compare the simulations with the HVC observations in the
MW, we set the position of the ``observer'' to an arbitrary position
on the disk plane of the simulated galaxies, with galactocentric
distance of $8.5 \kpc$, and ``observe'' the \HI{} column density and
velocity of the gas particles from that position.  \Fig{HI_mom0}
demonstrates the \HI{} column density of HVCs using all-sky Aitoff
projections.  Here, we define HVCs as consisting of gas particles with
absolute lines-of-sight velocities, \vlsr{}, deviating from the Local
Standard of Rest (LSR) by more than $100 \kms$---\Fig{HI_mom0} shows
the \HI{} column density of these gas particles with $|\vlsr| >
100\kms$.  We set the rotation velocity of the LSR to $220 \kms$,
similar to that of the MW \citet{lmp+02}.  We can confirm that both
simulated galaxies have gas disks rotating at $\sim 220 \kms$.  In
this paper, results are based only on those particles within two
virial radii (\rvir; see \Tab{params}).  We have also confirmed that
the results are not sensitive to the cutoff radius chosen for column
densities $N(\HI{}) \simgt 10^{17} \cm$.  We only display results for
one chosen observer, however we have confirmed the generality of these
results, with the sky coverage fraction typically changing by no more
than 20\% for a given column density, as we change the observer's
position and/or we analyze other outputs of the simulation near the
final redshift.

Our chemodynamical simulation follows the hydrogen and other elemental
abundances for each gas particle, but does not calculate the
ionization fraction of each species.  Instead, the \HI{} mass for each
gas particle is calculated assuming collisional ionization equilibrium
(CIE).  The CIE neutral hydrogen fraction is estimated using \cloudy94
\citep{fkv+98}.  We multiply the fraction by the hydrogen abundance,
and obtain a \HI{} mass fraction for each particle as a function of
its temperature.  We ignore any effect from the background radiation
field.

\section{Results and discussion}
\label{sec:results}

\Fig{HI_mom0} demonstrates that both galaxies have a significant
number of HVCs, with column densities comparable to those observed by
\citet{w91}.  KGCD displays several large linear HVCs at galactic
longitudes $l \sim 60\degr$ and $l \sim 270\degr$.  These components
correspond to the outer ring structure seen at a galactocentric radius
of $\sim 30 \kpc$ in \Fig{gasdens}.  We name this structure the
``polar gas ring'', and discuss it later.

\begin{figure}
  \epsscale{1.1}
  \plotone{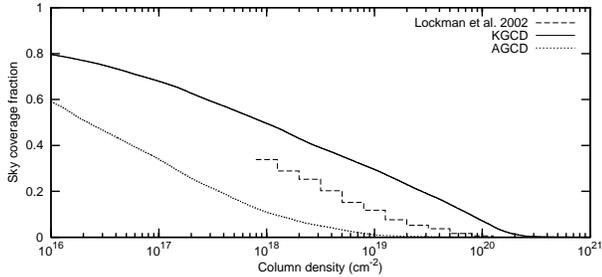}
  \caption[\HI{} sky coverage fraction vs.\ limiting column
  density]{\HI{} sky coverage fraction vs.\ limiting column
    density.  The dashed line denotes the sky coverage from the
    \citet{lmp+02} HVC survey (excluding velocities \vlsr{} in the
    range $\pm 100 \kms$, and with poor coverage of latitudes
    $|b|<20\degr$) to the $4\sigma$ completeness limit.  The sky
    coverage obtained from the KGCD simulation is shown as a solid
    line, and the dotted line denotes the AGCD simulation. }
  \label{fig:hist_HI_skycov}
\end{figure}

To compare the HVC population of our simulations with the MW HVCs
quantitatively, \Fig{hist_HI_skycov} shows the fraction of sky covered
by HVCs as a function of limiting column density for both simulations
and for the observations in \citet{lmp+02}.  In this plot, we exclude
the area with low galactic latitude $|b| < 20\degr$, to avoid
contamination by the disk component (the sample of sightlines in
\citealp{lmp+02} was limited in a similar fashion).  As is obvious
from \Fig{HI_mom0}, KGCD has more high column density HVCs than AGCD,
and almost all of the sky is covered down to $10^{16} \cm$.  At a
fixed column density, the sky coverage in the simulations bracket the
observed sky coverage in the MW.  Note that we ignore any effects of
background radiation.  It is expected that such a field would decrease
the population of HVCs with Galactocentric distance less than $10
\kpc$ or $N(\HI{}) \simlt 10^{19} \cm$ \citep{m93,bm97,bm99}.  Thus,
since as mentioned below, the distances of the simulated HVCs are
greater than $10 \kpc$, our estimated coverage fractions should be
interpreted as an upper limit for $N(\HI{}) \simlt 10^{19} \cm$, and a
lower limit for $N(\HI{}) \simgt 10^{19} \cm$, where the highest
column density HVCs may not be fully resolved.  With these caveats,
KGCD appears to have a sufficient population of HVCs to explain the
observed population of HVCs within the MW.  We conclude that current
cosmological simulations can produce MW-size disk galaxies with
similar populations of HVCs to those in the MW.  The differences
between KGCD and AGCD may demonstrate real differences in the
populations of HVCs among disk galaxies.  However, to understand the
causes of such differences, we need a larger sample of high-resolution
simulated disk galaxies, the subject of a future study.

\begin{figure}
  \epsscale{1.1}
  \plotone{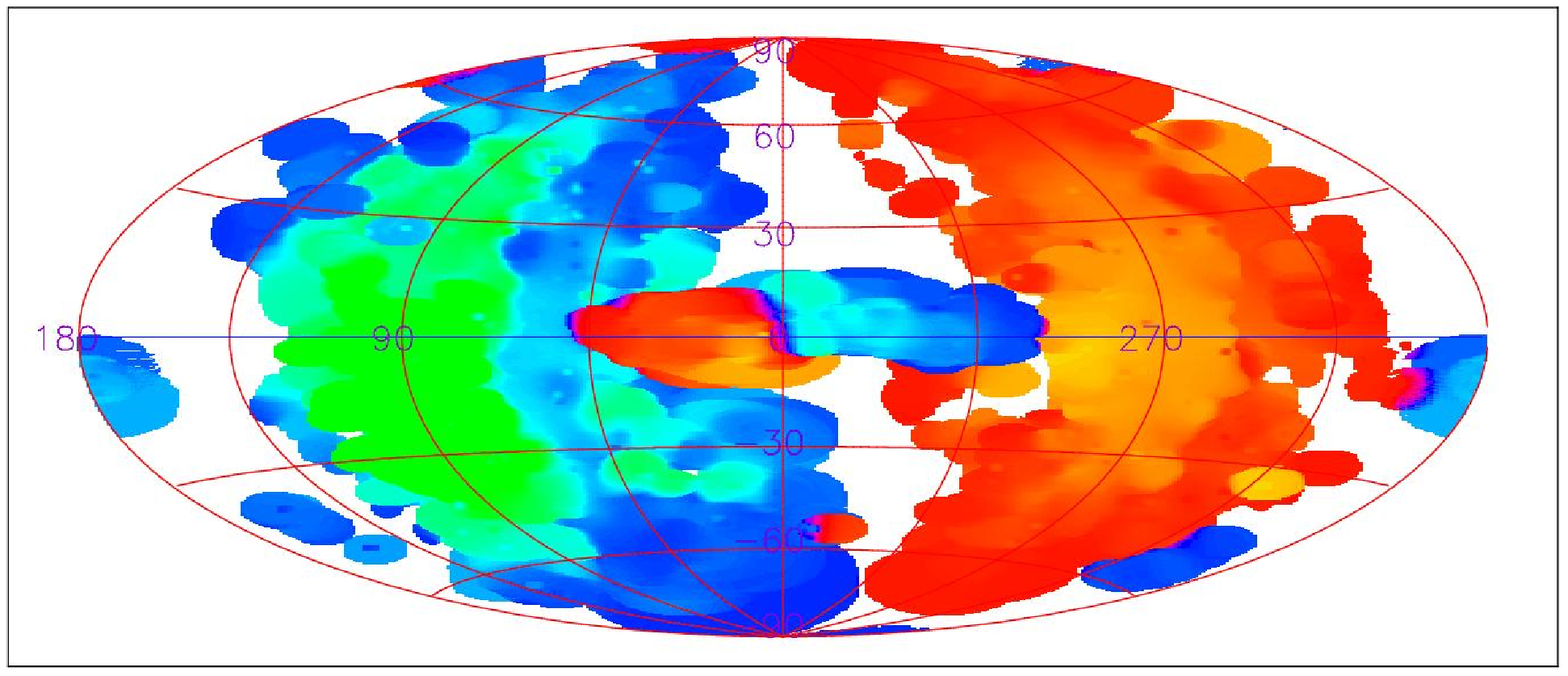}\\
  \plotone{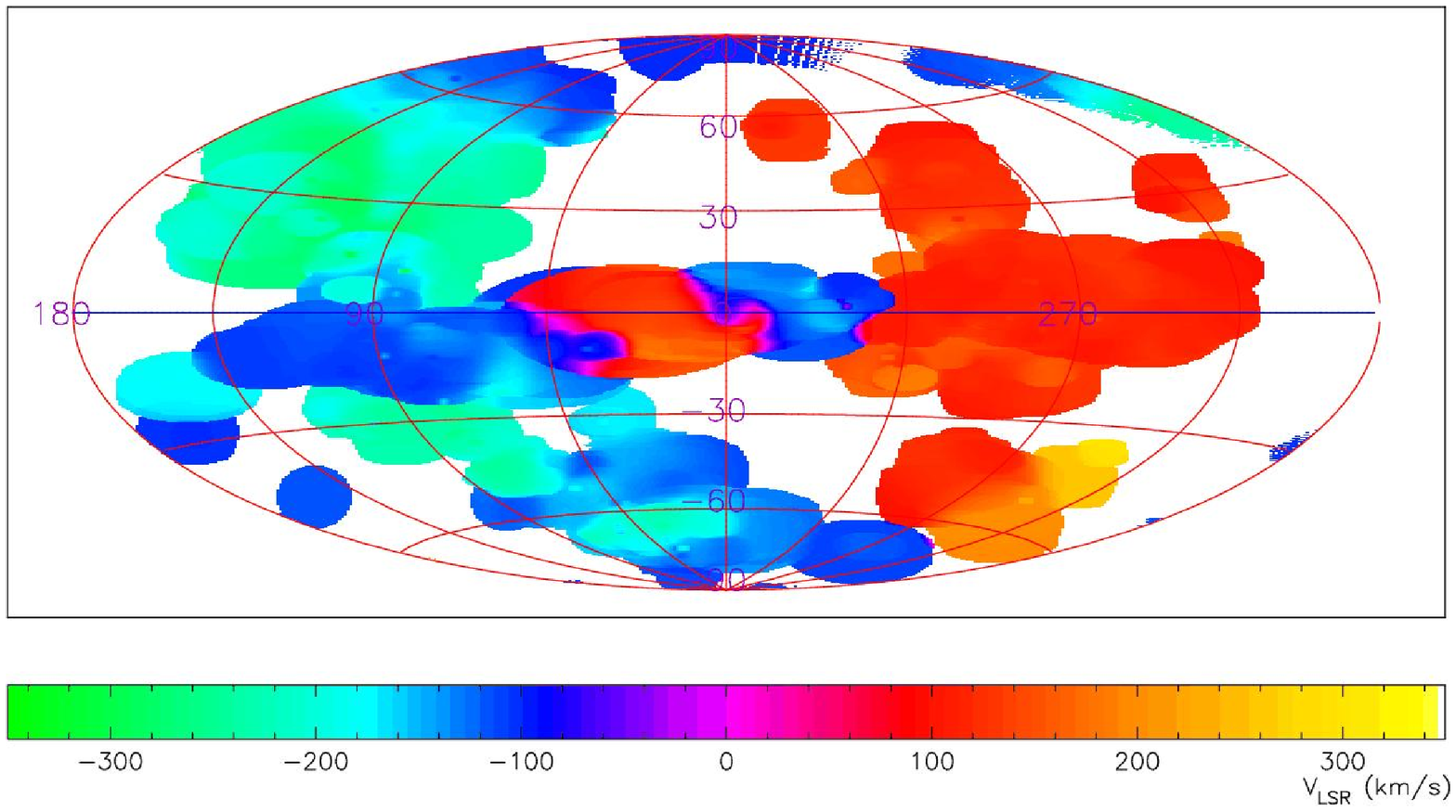}
  \caption[\HI{} velocity map]{\HI{} first moment (column density
    weighted mean velocity in the LSR) map of KGCD (top) and AGCD
    (bottom) simulations.  Particles with $|\vlsr| < 100\kms$ are
    excluded.  Compare these maps with fig.~16 in \citet{wss+03}.}
  \label{fig:HI_mom1}
\end{figure}

We find that velocities of the simulated HVCs seen in \Fig{HI_mom1},
are distributed similarly to those observed by \citet{pds+02}.
Overall, the clouds between $l=0\degr$ and $180\degr$ have a negative
velocity, while the clouds between $l=180\degr$ and $360\degr$ have a
positive velocity.  This is natural, since the LSR is moving towards
$l=90\degr$.  In KGCD, we find that there is a relatively large HVC
complex whose velocity is very high, $-450 \simlt \vlsr \simlt
-300\kms$.  These very high velocity clouds (VHVCs) are located
between $-45\degr \simlt b \simlt 15\degr$ and $60\degr \simlt l
\simlt 120\degr$, in the top panel of \Fig{HI_mom1}.  The MW also has
such VHVCs in the Anti-Center complex \citep{h78,hw88}.  The
galactocentric distance to the VHVC in KGCD is $\sim 10$--$25\kpc$,
and we find that the cloud is a gas clump which has recently fallen
into the galaxy.  We convert the \HI{} mass for each particle into a
\HI{} flux using $M_\HI = 0.235 D^2_{\rm kpc} S_{\rm tot}$
\citep{ww91}, where the \HI{} mass $M_\HI$ is in \msunns{}, total
\HI{} flux in \Jykmsns{} is $S_{\rm tot}$, and distance from observer
to the particle in \kpcns{} is $D_{\rm kpc}$.  We find the \HI{} mass
of the VHVC complex is $1.2\e8 \msun$, and the total \HI{} flux (for
the chosen observer) is $2.5\e6 \Jykms$.  Since it is
% see plots_of_interest: ./separatemovie.workerthread --masscolumn mzHgI --output columns noprint --outputfile /dev/null posselect.g000022.masses_hvcs/00\:chvc1.select.dat posselect.g000022.masses_hvcs/01\:chvc2.select.dat -- output/data/g000022.processed.OVI=min:0.rotdisk=090.vlim=100.wcstype=ait
infalling on a retrograde orbit, its velocity relative to the LSR
becomes very large depending on its location relative to the observer.
Therefore, the observed VHVCs may be explained by such infalling gas
clumps within the Galaxy.  It is also worth noting that we do not find
any associated stellar or DM components in the simulated VHVCs.

\begin{figure}
  \epsscale{1.1}
  \plotone{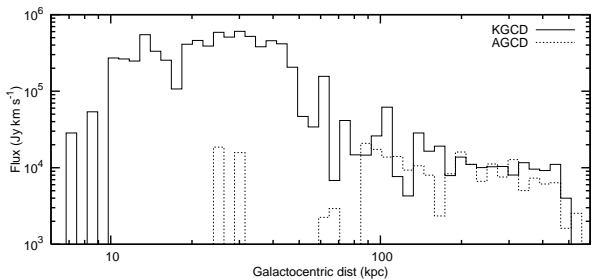}
  \caption[\HI{} distance histogram]{\HI{} flux weighted histogram
    with logarithmically spaced bins in distance (de-emphasing fluxes
    from smaller galactocentric distances) of both AGCD and KGCD
    simulations, excluding $|\vlsr{}| < 100 \kms$ and $|b|<20\degr$.}
  \label{fig:hist_HI_flux_dist}
\end{figure}

In the simulations, we are able to measure the distance of HVCs---data
which are not yet generally available for the real MW.  In
\Fig{hist_HI_flux_dist}, we show a flux-weighted histogram of the
galactocentric distances, \rmw{}, of the high velocity gas particles
with $|b| > 20\degr$.  In both simulations, less than 1\% of the \HI{}
flux visible in \Fig{HI_mom0} originates from HVCs with $\rmw
>150\kpc$.  This is consistent with the aforementioned limits
established by \citet{pbg+04}.

It is also clear that most of the emission in the KGCD simulation
results from the polar gas ring, whose radius is $\sim 30\kpc$.  The
mass of this ring (including low velocity gas) is $3.6\e9 \msun$, of
which $2.3\e9 \msun$ is \HI{}.  Although there is no prominent polar
% g000022.processed.rliml=20.rlimu=50.rotdisk=090.vlim=0.wcstype=ait
ring in AGCD, at larger distances, the flux distribution is found to
follow that of the KGCD simulation.

\Fig{HI_mom0} demonstrates that the polar gas ring forms a linear high
velocity structure found in all quadrants of the sky.  Large HVC
components, such as Complex~C and the Magellanic Stream, are well
known in our MW, and several authors \citep[e.g.][]{h88} argue that
the MW is surrounded by a polar gas ring.  Neglecting the Magellanic
Stream, which has likely originated from the infalling Magellanic
clouds \citep{gn96, yn03, ckg05}, the largest HVC structure is
Complex~C.  We measure the mass of the HVC ring in one quadrant as
$\sim 4\e8 \msun$.  If we place Complex~C at a galactocentric distance
of
%awk '{printf "%5.5e\n", j+=$35}' quad-0-50kpc.dat | tail -n 1
$30 \kpc$ (only a lower limit has been obtained for its galactocentric
distance of $8.8\kpc$; \citealp{wps+99a}), its mass would be $\sim\
1.5\e7 \msun$.  Nevertheless, this indicates that there is enough
\HI{} gas to create such large observed HVCs in the simulated galaxy.
In KGCD, the polar gas ring is a relatively recently formed structure
that begins forming at redshift $z \sim 0.2$, and prior to this time,
the associated gas particles are found flowing inward along
filamentary structures.  Thus, this simulated ring structure
demonstrates that current $\Lambda$CDM numerical simulations can
explain the existence of such large HVCs like Complex~C, as recently
accreted gas which rotates in a near-polar orbit.  On the other hand,
AGCD does not show such a prominent large structure.  This may
indicate that such HVC structures are not common in all disk galaxies.
Although this study focuses on only two simulated galaxies due to the
limit of our computational resources, more samples of high-resolution
simulated galaxies would elucidate how common such large structures
are, and what kind of evolution history is required to make such HVCs.

We also analyze the metallicity of the simulated HVCs.  In KGCD, we
find that the prominent HVCs have \HI{} flux weighted metallicities of
$-4 \simlt \log(Z/\Zsun) \simlt -2$ with a flux weighted mean of
$\log(Z/\Zsun) \sim -2.4$.  This is much lower than the metallicities
of the observed HVCs in the MW, $-1 < \log(Z/Z\sun) < 0$ \citep{w01}.
The polar gas ring seen in KGCD has a metallicity of $-3 \simlt
\log(Z/Z\sun) \simlt 0.5$ with a mean of $\log(Z/\Zsun) \sim -1.7$,
which is lower than the observed metallicity of Complex~C,
$\log(Z/Z\sun) \sim -1$ \citep[e.g.][]{ggp+01,rsw+01}.  Thus, our
numerical simulations seem to underestimate the metallicity of the
later infalling gas clouds.  This is likely because we adopt a weak
supernova feedback model in our simulations.  If we use a model with
strong feedback, more enriched gas is blown out from the system at
high redshift, which can enrich the inter-galactic medium which then
falls into galaxies at a later epoch.

The majority of the simulated HVCs, including the polar gas ring, do
not have any obviously associated stellar or DM components, which is
consistent with current observations \citep{sb02,smg+05}.  However, a
few compact HVCs are found to be associated with stellar components.
It would be interesting to estimate how bright they are, and if they
are detectable within the current observational limits.
Unfortunately, the resolution of the current simulations are too poor
to estimate the luminosity, and it is also likely that our simulations
produce too many stars due to our assumed minimal effect of supernova
feedback.

The clouds in the simulation may be destroyed by effects that our
simulations are not able to reproduce accurately.  The resolution and
nature of the SPH simulations make it difficult to resolve shocks
between the HVCs and the MW halo; however simple analytic estimates
show that there will not be strong shocking of the HVCs due to the low
density of the halo.  \citet{is04}, using more detailed simulations,
argue that the Mach number of the HVCs are only $M_s \sim 1.2$--$1.5$,
marginally sufficient to form shocks, and heating the leading $\sim
0.1 \pc$ of the HVC.  \citet{qm01} show that the lifetime of such HVCs
in the presence of shocks is $\sim 1\Gyr$, long enough for our HVCs to
survive.  \citet{mb04} discuss various physical processes, including
conduction, evaporation, ram-pressure drag, Jeans instabilities, and
Kelvin--Helmholtz instabilities that limit the mass ranges of stable
clouds.  The most massive HVC apparent in our simulations is found to
be $1.5\e8\msun$, and the mass
%./separatemovie.workerthread --masscolumn mass --output columns noprint --outputfile /dev/null posselect.g000022.masses_hvcs/00\:chvc1.select.dat posselect.g000022.masses_hvcs/01\:chvc2.select.dat -- output/data/g000022.processed.OVI=min:0.rotdisk=090.vlim=100.wcstype=ait
resolution and hence smallest resolvable HVC in our simulations are
$\sim 10^6 \msun$; clouds of both extremes are clearly within the
stable mass ranges summarized in fig.~6 of \citet{mb04}.  Therefore,
the simulated clouds are also expected to be stable.

We report that HVCs seem to be a natural occurrence in a $\Lambda$CDM
Universe.  We emphasize that the galaxies that result from our
simulations were not created specifically to reproduce the MW
exactly---they were selected for resimulation at higher resolution on
the basis of being disk-like \lstar{} galaxies.  However, we have
serendipitously discovered that simulated galaxies that are similar in
size to the MW naturally contain \HI{} gas in the vicinity of the disk
that are similar to the anomalous velocity features seen in the MW.

\acknowledgments

TC thanks Stuart Gill for his \pview{} visualization tool source code
and Chris Power for useful advice.  We acknowledge the Astronomical
Data Analysis Center of the National Astronomical Observatory, Japan
(project ID: wmn14a), the Institute of Space and Astronautical Science
of Japan Aerospace Exploration Agency, and the Australian and
Victorian Partnerships for Advanced Computing, where the numerical
computations for this paper were performed.  DK acknowledges the
financial support of the JSPS, through the Postdoctoral Fellowship for
research abroad.  The financial support of the Australian Research
Council and the Particle Physics and Astronomy Research Council of the
United Kingdom is gratefully acknowledged.

\bibliographystyle{apj}

%\bibliography{ms}

\end{document}